# Upgrades and maintenance of the CRYRING@ESR electron cooler for improved internal electron target operation

A. Koutsostathis[1,*], C. Krantz[2], E.-O. Hanu[2], F. Herfurth[2], R. Heß[2],
W. Kaufmann[2], M. Lestinsky[2], M. Looshorn[3,4], E. B. Menz[2], A. Reiter[2],
J. Roßbach[2], S. Schippers[3,4], P. M. Suherman[2]

[1] *Department of Physics, National Technical University of Athens, 15780 Zografou, Greece*
[2] *GSI Helmholtzzentrum für Schwerionenforschung, 64291 Darmstadt, Germany*
[3] *I. Physikalisches Institut, Universität Gießen, 35392 Giessen, Germany*
[4] *Helmholtz Forschungsakademie Hessen für FAIR (HFHF), Campus Gießen,
GSI Helmholtzzentrum für Schwerionenforschung, 35392 Giessen, Germany*

**Abstract**  The electron cooler of the CRYRING@ESR storage ring at the GSI-FAIR accelerator complex is a unique instrument, which not only provides beam cooling of the stored ions, but also serves as a low-energy electron target for Dielectronic Recombination experiments. The minimisation of vacuum contamination and the response to rapid energy changes are key requirements of the cooler in electron target mode. Therefore, a test bench was prepared to study the outgassing behaviour of the electron gun components and a setup was constructed to evaluate the drift-electrode-modulation of the acceleration voltage of the cooler. The vacuum studies showed that the electron gun cathode was severely malfunctioning and resulted in its replacement. The drift-electrode-modulation of the acceleration voltage showed significant improvements compared to direct modulation of the terminal voltage of the cooler.

**Keywords**  Electron cooler, storage ring, dielectronic recombination, electron gun

## INTRODUCTION

CRYRING@ESR is a low-energy heavy-ion storage ring at the GSI/FAIR accelerator complex, in which decelerated ions from the upstream accelerator chain or from a local injector are stored and cooled for experimental usage. Originally located at the Manne Siegbahn Laboratory in Stockholm, it was relocated to GSI/FAIR, installed downstream from the ESR, and, following a number of upgrades, has been providing beam for experiments in the FAIR Phase 0 programme since 2021 [1].

*The electron cooler of CRYRING@ESR*

The electron cooler of CRYRING@ESR improves the quality of the stored ion beams and, additionally, serves as an electron target for Dielectronic Recombination (DR) experiments. In electron cooling, a dense cold electron beam interacts with an ion beam, with both beams overlapping coaxially and at equal average velocity. Due to the large mass difference between electrons and ions, the heat exchange between the two species results in compression of the ion beam in phase space [2].

The components of the electron cooler are shown in Fig. 1. The electron beam, which is produced via thermionic emission from a BaO dispenser cathode, is adiabatically expanded and then directed to the interaction region, where the cooling process takes place [3]. The average velocity of the electrons is primarily defined by the difference in electrical potential between the emitting cathode and the interaction region. Finally, the electron beam is dumped in the collector of the cooler [4].

*Dielectronic Recombination Experiments*

DR involves the resonant capture of an electron by an atomic ion via an intermediate auto-ionising state and is the dominant recombination process in various plasma environments. Electron coolers are ideal platforms for precision measurements of DR if the electron velocity is shifted with respect to that of the ions, making the cooler an electron target [5]. For quasi-simultaneous cooler and target operation, the electron energy is cycled between the "matching" and "probing" conditions by application of a rapid detuning voltage ($U_d$) on top of the bias voltage ($U_c$) of the electron gun. Recombined ions have different mass-to-charge ratio compared to the stored beam, and the next dipole magnet directs them towards a scintillation counter (cf. Fig. 2) [1].

---

* Corresponding author: athanasios.koutsostathis@cern.ch

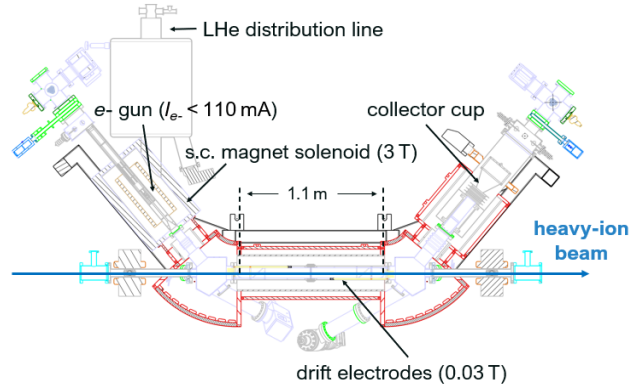

**Figure 1**. *Schematic of the electron cooler of CRYRING@ESR.*

Limiting factors in DR experiments include the residual gas pressure in the cooler section and instabilities of the electron beam energy, and therefore, collision energy. The former leads to high detector background, as ions capture electrons from residual gas molecules in addition to DR. The latter can arise from insufficient response to the fast detuning voltage applied in the experiment, blurring the energy resolution.

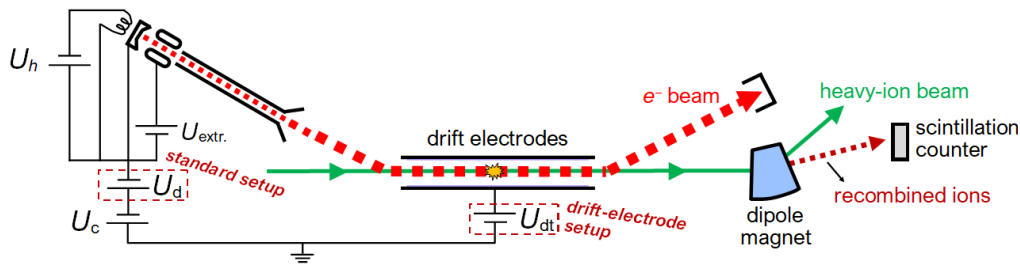

**Figure 2**. *Simplified schematic of the principles of the DR experiment setup at CRYRING@ESR and of the standard and drift-electrode setups, which define the electron beam energy.*

## ELECTRON GUN VACUUM TESTS

Vacuum contamination caused by the outgassing of the electron gun's cathode is a potential source of detector background for the DR experiments. During the latest beam time period (01.-06.2024) a heating power of approximately 14 V×1.3 A was needed to reach the required cathode temperature of approximately 900 °C, as opposed to the expected power of approximately 9 V×1 A. Furthermore, the visible glow of the hot cathode was inhomogeneous, suggesting uneven temperature distribution. The high power is likely to have influenced the outgassing of the gun's components.

To better understand these impairments, the electron gun was removed from the cooler and placed into a test setup at the GSI/FAIR vacuum lab. The setup encompassed a vacuum chamber equipped with ionisation pressure gauges, a residual gas analyser, electrical feedthroughs for contacting the gun's electrodes and cathode heating, and a view port through which the temperature of the cathode could be measured with a pyrometer. After bakeout at approximately 200 °C and evacuation by a turbo pump, the test chamber reached a base pressure of the order of $10^{-10}$ mbar.

The cathode filament was heated by application of a standard heating voltage ($U_h$) of 8.5 V to 9.0 V to the heating filament, however reaching only 600 °C to 700 °C of measured temperature at the cathode surface. During this process, a significant rise in the partial pressures of $H_2$ and $H_2O$ was observed (cf. Fig. 3), which was attributed to air exposure during transportation, and had been anticipated. After overnight degassing of the water contamination, the aforementioned abnormally high heating power was applied to reach a measured temperature of 800 °C to 900 °C. Even so, only a very weak emission current of the order a few tens of microamperes was measured at an applied extraction voltage of 500 V. At the high heating power, significant outgassing of CO and $CO_2$ was observed (cf. Fig. 3) which is a possible source for the high background in DR measurements.

Based on these tests, it was concluded that the cathode was severely malfunctioning, possibly due to surface poisoning, and it was replaced. Additionally, heavily metallised surfaces on the gun isolators, which were formed due to the abnormally-high-power operation, were replaced and non-vented volumes in the electron gun were fixed such that they could be evacuated. Following evacuation and bakeout, a new cathode was briefly heated to approximately 850 °C, where it displayed a much more homogeneous glow of the hot surface. Further studies after the activation of the cathode inside the CRYRING@ESR vacuum have confirmed these observations. After an improved bakeout and outgassing procedure, and operating at comparatively lower power, the new cathode shows promising behaviour for the upcoming experimental runs.

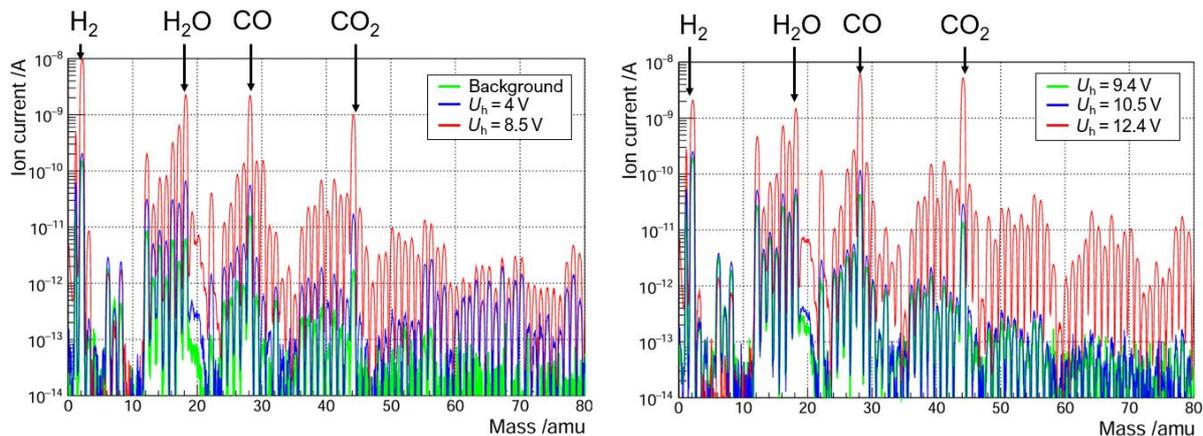

**Figure 3**. *Residual gas mass spectra in the test setup before, during and after heating of the cathode to the anticipated operation conditions (left) and, following overnight degassing, during further heating up to 900 °C (right). The spectrum lines of characteristic vacuum contaminants ($H_2$, $H_2O$, CO and $CO_2$) are labelled.*

## STEP RESPONSE FOR ELECTRON TARGET OPERATION

Application of the energy-changing detuning voltage to the potential of the electron gun has caused problems in past DR experiments. In the standard setup $U_d$ is provided by a four-quadrant high-voltage (HV) amplifier (KEPCO BOP, ± 1000 V), while the cathode bias $U_c$ is provided by a high-precision, yet slow, unipolar supply [1]. In hitherto standard operation, both voltage supplies are connected in series (cf. Fig. 2) to define the gun potential. In this configuration repeated rapid changes of $U_d$ on the millisecond-scale lead to alternating high currents which can cause instability of the control circuit of $U_c$ and therefore of the overall acceleration potential, $U_{acc} = -U_c - U_d$ (cf. Fig. 4).

The recent installation of drift-tube electrodes in the merged-beams' region will allow for local variation of the electron energy, separating the HV supply into different sub-systems (cf. Fig. 2), thus avoiding interference and reducing the overall capacitance of the oscillating system. In order to test the step-response of the drift-electrode potential, $U_{dt}$, the electrodes were connected to the fast HV amplifier, a rectangular function was applied to the amplifier's input, and the response of the electrodes' potential to sharp voltage steps, of amplitudes of several hundreds of volts, was observed with a high-bandwidth (40 MHz) HV probe.

The drift-electrode setup significantly improves the behaviour of the oscillating acceleration potential, $U_{dt}$, which allows for better approximations of rectangular-step-like changes of $U'_{acc} = -U_c + U_{dt}$. It was observed that the 10%-to-90% rise time, which is independent of the voltage jump's amplitude, improved from approximately 0.7 ms to 0.2 ms. Some residual time dependence of the signal still remains on the millisecond scale, as shown in Fig. 4, but is observed identically with the drift electrodes being disconnected, suggesting that the bandwidth of the HV amplifier itself is now the limiting factor in the experiment.

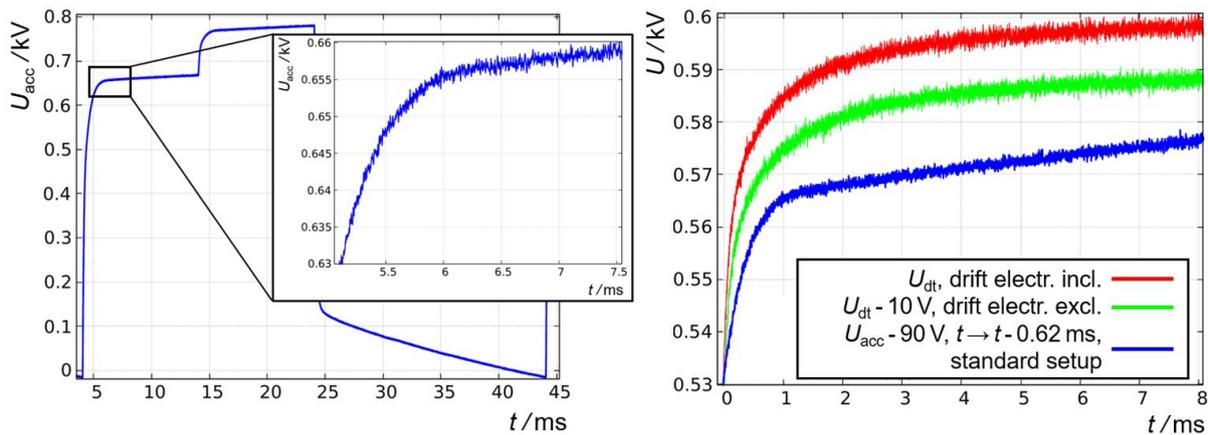

*Figure 4.* Left: Step response of the sum acceleration voltage ($U_{acc}$) for the standard energy modulation scheme, with interference between the two voltage supplies leading to the instability of $U_{acc}$. Right: Measurements of the alternating voltage, $U_{dt}$, applied to the drift electrodes (red curve) show a higher slew rate following step-wise change of the set-point. The green curve (offset by -10 V) shows the behaviour of the alternating voltage, $U_{dt}$, with all loads being disconnected.

## CONCLUSION AND PROSPECTS

The electron cooler is a key element of the CRYRING@ESR storage ring, both, as a means of beam preparation through cooling and as an experimental platform for DR measurements. Using a vacuum test setup, a better understanding of the outgassing behaviour of the hot cathode of the electron gun has been reached; its resulting replacement has improved the overall operation of the electron gun. Furthermore, initial tests of the new drift-tube setup for energy modulation revealed it to be a promising option for further improvement of the energy resolution in DR experiments.

**Acknowledgments**

A.K. would like to acknowledge the Helmholtz Graduate School for Hadron and Ion Research, for the organisation and funding of the International Summer Student Programme at GSI-FAIR, and especially Dr. Ralf Averbeck and Ms. Gabriela Menge, for the management of the programme. Financial support from the German Federal Ministry for Education and Research (BMBF) via the Collaborative Research Centre ErUM-FSP T05 – "Aufbau von APPA bei FAIR" (Grant Nos. 05P19RGFA1, and 05P21RGFA1) is also gratefully acknowledged.